\documentclass{article}

\usepackage{microtype}
\usepackage{graphicx}
\usepackage{subfigure}
\usepackage{booktabs} 
\usepackage[table]{xcolor}
\usepackage{graphicx}
\usepackage{textcomp}
\usepackage{natbib}
\usepackage{multirow}

\usepackage{hyperref}

\usepackage[accepted]{icml2020}

\icmltitlerunning{Predicting COVID-19 Pneumonia Severity on Chest X-ray with Deep Learning}

\begin{document}
\setcitestyle{square}
\twocolumn[
\icmltitle{Predicting COVID-19 Pneumonia Severity on Chest X-ray with Deep Learning}



\icmlsetsymbol{equal}{*}

\begin{icmlauthorlist}
\icmlauthor{Joseph Paul Cohen}{mila,udem}
\icmlauthor{Lan Dao}{mila,fm}
\icmlauthor{Paul Morrison}{mila,f}
\icmlauthor{Karsten Roth}{mila,hei}
\icmlauthor{Yoshua Bengio}{mila,udem}
\icmlauthor{Beiyi Shen, MD}{sb}
\icmlauthor{Almas Abbasi, MD}{sb}
\icmlauthor{Mahsa Hoshmand-Kochi, MD}{sb}
\icmlauthor{Marzyeh Ghassemi}{vector,uot}
\icmlauthor{Haifang Li}{sb}
\icmlauthor{Tim Q Duong}{sb}


\end{icmlauthorlist}

\icmlaffiliation{mila}{Mila, Quebec Artificial Intelligence Institute}
\icmlaffiliation{vector}{Vector Institute}
\icmlaffiliation{hei}{Heidelberg University}
\icmlaffiliation{udem}{University of Montreal}
\icmlaffiliation{uot}{University of Toronto}
\icmlaffiliation{fm}{Faculty of Medicine, University of Montreal}
\icmlaffiliation{f}{Fontbonne University}
\icmlaffiliation{sb}{Stony Brook Medicine}

\icmlcorrespondingauthor{Joseph Paul Cohen}{joseph@josephpcohen.com}

\icmlkeywords{Computer Vision, COVID-19}

\vskip 0.3in
]



\printAffiliationsAndNotice{}  

\begin{abstract}
\textbf{Purpose:} The need to streamline patient management for COVID-19 has become more pressing than ever. Chest X-rays provide a non-invasive (potentially bedside) tool to monitor the progression of the disease. In this study, we present a severity score prediction model for COVID-19 pneumonia for frontal chest X-ray images. Such a tool can gauge severity of COVID-19 lung infections (and pneumonia in general) that can be used for escalation or de-escalation of care as well as monitoring treatment efficacy, especially in the ICU. 

\textbf{Methods:} Images from a public COVID-19 database were scored retrospectively by three blinded experts in terms of the extent of lung involvement as well as the degree of opacity. A neural network model that was pre-trained on large (non-COVID-19) chest X-ray datasets is used to construct features for COVID-19 images which are predictive for our task. 

\textbf{Results:} This study finds that training a regression model on a subset of the outputs from an this pre-trained chest X-ray model predicts our geographic extent score (range 0-8) with 1.14 mean absolute error (MAE) and our lung opacity score (range 0-6) with 0.78 MAE. 

\textbf{Conclusions:} These results indicate that our model’s ability to gauge severity of COVID-19 lung infections could be used for escalation or de-escalation of care as well as monitoring treatment efficacy, especially in the intensive care unit (ICU). A proper clinical trial is needed to evaluate efficacy. To enable this we make our code, labels, and data available online  at \href{https://github.com/mlmed/torchxrayvision/tree/master/scripts/covid-severity}{GitHub:covid-severity} and \href{https://github.com/ieee8023/covid-chestxray-dataset}{GitHub:covid-chestxray-dataset}

\end{abstract}

\section{Introduction}

As the first countries explore deconfinement strategies \cite{Wilson2020} the death toll of COVID-19 keeps rising \cite{OGrady2020}. The increased strain caused by the pandemic on healthcare systems worldwide has prompted many physicians to resort to new strategies and technologies. Chest X-rays (CXRs) provide a non-invasive (potentially bedside) tool to monitor the progression of the disease \cite{Yoon2020,Ng2020}. As early as March 2020, Chinese hospitals used artificial intelligence (AI)-assisted computed tomography (CT) imaging analysis to screen COVID-19 cases and streamline diagnosis \cite{Jin2020}. Many teams have since launched AI initiatives to improve triaging of COVID-19 patients (i.e., discharge, general admission or ICU care) and allocation of hospital resources (i.e., direct non-invasive ventilation to invasive ventilation) \cite{Strickland2020}. While these recent tools exploit clinical data, practically deployable CXR-based predictive models remain lacking. 

In this work, we build and study a model which predicts the severity of COVID-19 pneumonia, based on CXRs, to be used as an assistive tool when managing patient care. The ability to gauge severity of COVID-19 lung infections can be used for escalation or de-escalation of care, especially in the ICU. An automated tool can be applied to patients over time to objectively and quantitatively track disease progression and treatment response. 


\section{Materials and Methods}

\subsection{COVID-19 Cohort}
We used a retrospective cohort of 94 posteroanterior (PA) CXR images from a public COVID-19 image data collection \cite{Cohen2020coviddataset}. While the dataset currently contains 153 images, it only counted 94 images at the time of the experiment, all of which were included in the study. All patients were reported COVID-19 positive and sourced from many hospitals around the world from December 2019 to March 2020. The images were de-identified prior to our use and there was no missing data. The ratio between male/female was 44/36 with an average age of 56$\pm$14.8 (55$\pm$15.6 for male and 57$\pm$13.9 for female).

\vspace{-5pt}
\subsection{Labels}
Radiological scoring was performed by three blinded experts: two chest radiologists (each with at least 20 years of experience) and a radiology resident. They staged disease severity using a score system adapted from \cite{10.1148/radiol.2020201160}, based on two types of scores (parameters): extent of lung involvement and degree of opacity. 

\vspace{-5pt}
\begin{enumerate}
    \itemsep0em 
    \item The extent of lung involvement by ground glass opacity or consolidation for each lung (right lung and left lung separately) was scored as: 0 = no involvement; 1 = $<$25\% involvement; 2 = 25-50\% involvement; 3 = 50-75\% involvement; 4 = $>$75\% involvement. The total extent score ranged from 0 to 8 (right lung and left lung together). 
    
    \item The degree of opacity for each lung (right lung and left lung separately) was scored as: 0 = no opacity; 1 = ground glass opacity; 2 = consolidation; 3 = white-out. The total opacity score ranged from 0 to 6 (right lung and left lung together). 
\end{enumerate}
\vspace{-5pt}
A spreadsheet was maintained to pair filenames with their respective scores. Fleiss’ Kappa for inter-rater agreement was 0.45 for the opacity score and 0.71 for the extent score..

\vspace{-5pt}
\subsection{Non-COVID-19 (Pre-Training) Datasets}

\begin{figure*}
    \centering
    \includegraphics[width=1\textwidth]{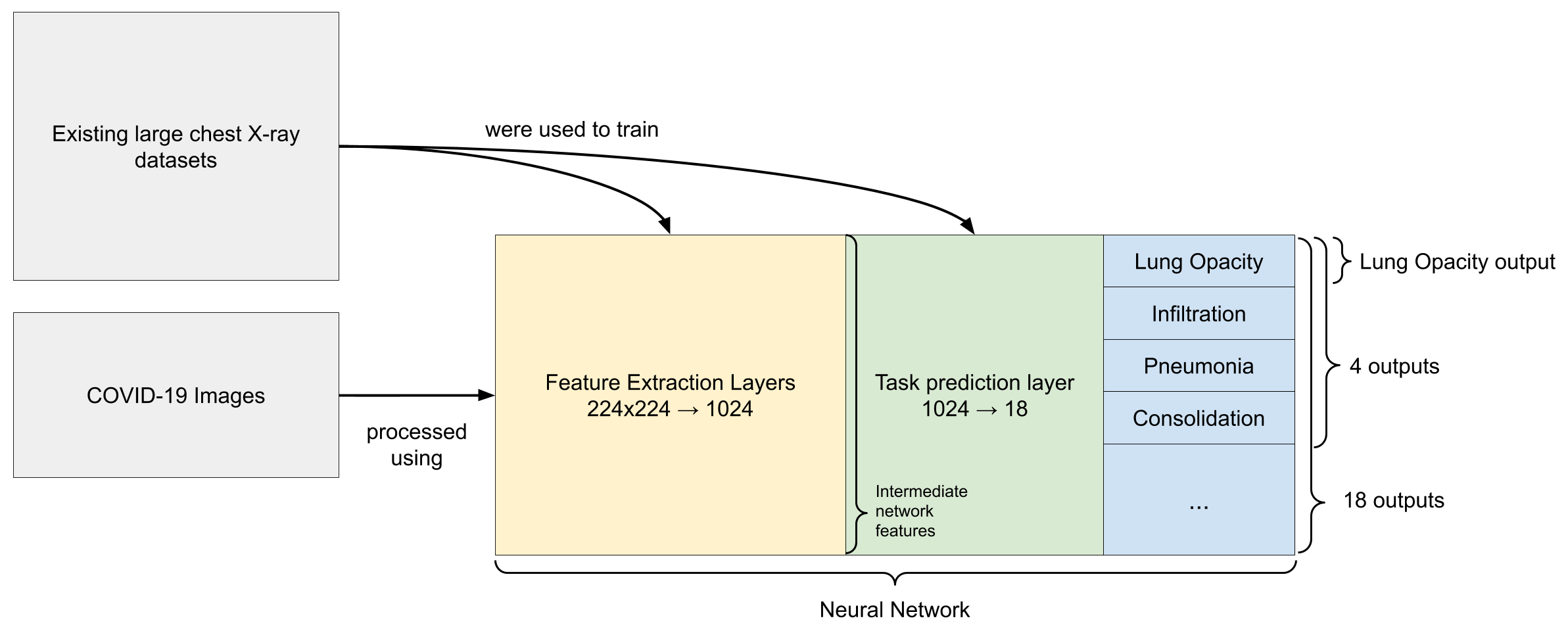}
    \vspace{-10pt}
    \caption{Detail of the different features being used. The two dataset blocks show that COVID-19 images were not used to train the neural network. The network diagram is split into 3 sections. The feature extraction layers are convolutional layers which transform the image into a 1024 dimensional vector which is called the intermediate network features. These features are then transformed using the task prediction layer (a sigmoid function for each task) into the outputs for each task. The different groupings of outputs used in this work are shown.}
    \label{fig:features}
\end{figure*}

Prior to the experiment, the model was trained on the following public datasets, none of which contained COVID-19 cases:

\vspace{-5pt}
\begin{enumerate}
    \itemsep0em 
    \item RSNA Pneumonia Challenge \cite{Shih2019RSNAKaggle};
    \item CheXpert - Stanford University\cite{Irvin2019CheXpert};
    \item ChestX-ray8 - National Institutes of Health (NIH) \cite{WangNIH2017};
    \item ChestX-ray8 - NIH with labels from Google \citep{Majkowska2019};
    \item MIMIC-CXR - MIT \cite{Johnson2019mimic-cxr};
    \item PadChest - University of Alicante \cite{Bustos2019PadChest};
    \item OpenI \cite{Demner-Fushman2016}
\end{enumerate}
\vspace{-5pt}
These seven datasets were manually aligned to each other on 18 common radiological finding tasks in order to train a model using all datasets at once (atelectasis, consolidation, infiltration, pneumothorax, edema, emphysema, fibrosis, fibrosis, effusion, pneumonia, pleural thickening, cardiomegaly, nodule, mass, hernia, lung lesion, fracture, lung opacity, and enlarged cardiomediastinum). For example ``pleural effusion" from one dataset is considered the same as ``effusion" from another dataset in order to consider these labels equal. In total, 88,079 non-COVID-19 images were used to train the model on these tasks.

\vspace{-5pt}
\subsection{Model, Preprocessing, and Pre-Training}
\vspace{-3pt}
In this study, we used a DenseNet model \cite{Huang2017} from the TorchXRayVision library \cite{Cohen2020xrv, cohen2020limits}. DenseNet models have been shown to predict Pneumonia well \cite{Rajpurkar2017chexnet}. Images were resized to 224 × 224 pixels, utilizing a center crop if the aspect ratio was uneven, and the pixel values were scaled to [-1024, 1024] for the training. More details about the training can be found in \cite{cohen2020limits}.

Before even processing the COVID-19 images, a pre-training step was performed using the seven datasets to train feature extraction layers and a task prediction layer (shown in Figure \ref{fig:features}). This ``pre-training" step was performed on a large set of data in order to construct general representations about lungs and other aspects of CXRs that we would have been unable to achieve on the small set of COVID-19 images available. Some of these representations are expected to be relevant to our downstream tasks. There are a few ways we can extract useful features from the pre-trained model as detailed in Figure \ref{fig:features}. 

\vspace{-5pt}
\subsection{Training}
\vspace{-3pt}
Similarly to the images from non-COVID-19 datasets used for pre-training, each image from the COVID-19 dataset was preprocessed (resized, centercropped, rescaled), then processed by the feature extraction layers and the task prediction layer of the network. The network was trained on existing datasets before the weights were frozen. COVID-19 images were processed by the network to generate features used in place of the images. As was the case with images from the seven non-COVID-19 datasets, the feature extraction layers produced a representation of the 94 COVID-19 images using a 1024 dimensional vector, then the fully connected task prediction layer produced outputs for each of the 18 original tasks. We build models on the pre-sigmoid outputs. 

Linear regression was performed to predict the aforementioned scores (extent of lung involvement and opacity) using these different sets of features in place of the image itself:
\vspace{-5pt}
\begin{enumerate}
    \itemsep-0.2em 
    \item Intermediate network features - the result of the convolutional layers applied to the image resulting in a 1024 dimensional vector which is passed to the task prediction layer;
    \item 18 outputs - each image was represented by the 18 outputs (pre-sigmoid) from the pre-trained model;
    \item 4 outputs - a hand picked subset of outputs (pre-sigmoid) were used containing radiological findings more frequent in pneumonia (lung opacity, pneumonia, infiltration, and consolidation); 
    \item Lung opacity output - the single output (pre-sigmoid) for lung opacity was used because it was task related. This is different from the predicted opacity score.
\end{enumerate}
\vspace{-15pt}

For each experiment performed, the 94 images COVID-19 dataset was randomly split into a train and test set roughly 50/50. Multiple timepoints from the same patient were grouped together into the same split so that a patient did not span both sets. Sampling was repeated throughout training in order to obtain a mean and standard deviation for each performance. As linear regression was used, there was no early stopping that had to be done to prevent the model from overfitting. Therefore, the criterion for determining the final model was only the mean squared error (MSE) on the training set.

\vspace{-5pt}
\subsection{Saliency maps}

In order to ensure that the models are looking at reasonable aspects of the images \cite{Reed1999smithing, Zech2018,Viviano2019}, a saliency map is computed by computing the gradient of the output prediction with respect to the input image (if a pixel is changed how much will it change the prediction). In order to smooth out the saliency map, it is blurred using a 5x5 Gaussian kernel. Keep in mind that these saliency maps have limitations and only offer a restricted view into why a model made a prediction \cite{Ross2017rrr,Viviano2019}.

\begin{table*}[t]
\centering
\caption{Performance metrics of each set of features for the Geographic Extent and Opacity Score predictions. Evaluation is performed on 50 randomly chosen train test splits and the metrics here are computed on a hold out test set. $R^2$ : coefficient of determination; MAE: mean absolute error; MSE: mean squared error. ``4 outputs" refers to lung opacity, pneumonia, infiltration, and consolidation.}
\label{tab:extentandopacity}
\begin{tabular}{llccccc}
\toprule
Task                               & Using features:               & \begin{tabular}[c]{@{}l@{}}\# parameters\\(fewer is better)\end{tabular} & \begin{tabular}[c]{@{}l@{}}Pearson\\Correlation\end{tabular} & $R^2$      & MAE       & MSE       \\
\midrule
\multirow{5}{*}{\begin{tabular}[c]{@{}l@{}}Opacity\\Score\end{tabular}}     & "lung opacity" output         & 1+1                             & \textbf{0.78$\pm$0.04}                                                     & \textbf{0.58$\pm$0.09}  & \textbf{0.78$\pm$0.05} & \textbf{0.86$\pm$0.11} \\
                                   & 4 outputs                     & 4+1                             & 0.78$\pm$0.04                                                     & 0.58$\pm$0.09  & 0.76$\pm$0.05 & 0.87$\pm$0.12 \\
                                   & 18 outputs                    & 18+1                            & 0.73$\pm$0.09                                                     & 0.44$\pm$0.16  & 0.86$\pm$0.11 & 1.15$\pm$0.33 \\
                                   & Intermediate network features & 1024+1                          & 0.66$\pm$0.08                                                     & 0.25$\pm$0.21  & 1.01$\pm$0.09 & 1.54$\pm$0.28 \\
                                   & No data                       & 0+1                             & 0.00$\pm$0.00                                                    & -0.08$\pm$0.10 & 1.24$\pm$0.10 & 2.26$\pm$0.36 \\
\midrule
\multirow{5}{*}{\begin{tabular}[c]{@{}l@{}}Geographic\\Extent\end{tabular}} & "lung opacity" output         & 1+1                             & \textbf{0.80$\pm$0.05}                                                     & \textbf{0.60$\pm$0.09}  & \textbf{1.14$\pm$0.11} & \textbf{2.06$\pm$0.34} \\
                                   & 4 outputs                     & 4+1                             & 0.79$\pm$0.05                                                     & 0.57$\pm$0.10  & 1.19$\pm$0.11 & 2.17$\pm$0.37 \\
                                   & 18 outputs                    & 18+1                            & 0.76$\pm$0.08                                                     & 0.47$\pm$0.16  & 1.32$\pm$0.17 & 2.73$\pm$0.89 \\
                                   & Intermediate network features & 1024+1                          & 0.74$\pm$0.08                                                     & 0.43$\pm$0.16  & 1.36$\pm$0.13 & 2.88$\pm$0.58 \\
                                   & No data                       & 0+1                             & 0.00$\pm$0.00                                                     & -0.08$\pm$0.10 & 2.00$\pm$0.17 & 5.60$\pm$0.95 \\
\bottomrule
\end{tabular}
\end{table*}

\begin{figure*}[!t]
\centering

    \includegraphics[width=0.5\textwidth]{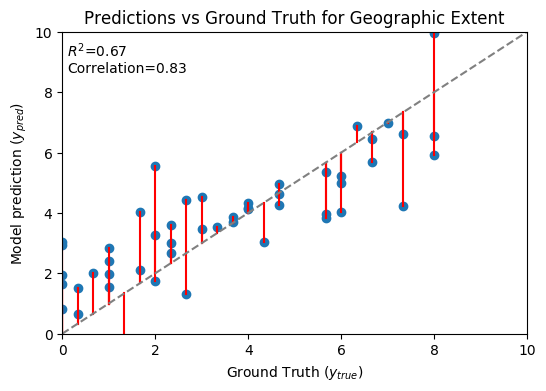}%
    \includegraphics[width=0.5\textwidth]{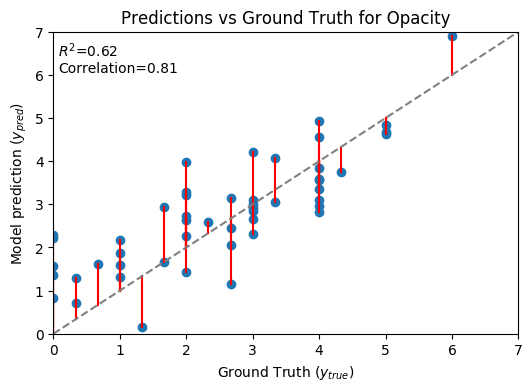}
    \vspace{-10pt}
    \caption{Scatter plots showing alignment between our best model predictions and human annotation (ground truth) for Geographical Extent and Opacity scores. Evaluation is on a hold out test set. The grey dashed line is a perfect prediction. Red lines indicate error from a perfect prediction. $R^2$ : coefficient of determination.}
    \label{fig:regression}
\end{figure*}

\section{Results}

\textbf{Quantitative performance metrics} The single ``lung opacity" output as a feature yielded the best correlation (0.80), followed by 4 outputs (lung opacity, pneumonia, infiltration, and consolidation) parameters (0.79) (Table \ref{tab:extentandopacity}). Building a model on only a few outputs provides the best performance. The mean absolute error (MAE) is useful to understand the error in units of the scores that are predicted while the mean squared error (MSE) helps to rank the different methods based on their furthest outliers. One possible reason that fewer features work best is that having fewer parameters prevents overfitting. Some features could serve as proxy variables to confounding attributes such as sex or age and preventing these features from being used prevents the distraction from hurting generalization performance. Hand selecting feature subsets which are intuitively related to this task imparts domain knowledge as a bias on the model which improves performance. Thus, the top performing model (using the single ``lung opacity" output as a feature) is used for the subsequent qualitative analysis.

\textbf{Qualitative analysis of predicted scores} Figure \ref{fig:regression} shows the top performing model's (using the single ``lung opacity" output as a feature) predictions against the ground truth score (given by the blinded experts) on held out test data. Majority of the data points fall close to the line of unity. The model overestimates scores between 1 and 3 and underestimates scores above 4. However, generally the predictions seem reasonable given the agreement of the raters.

\begin{figure*}[]
    \centering
    \includegraphics[width=0.9\textwidth]{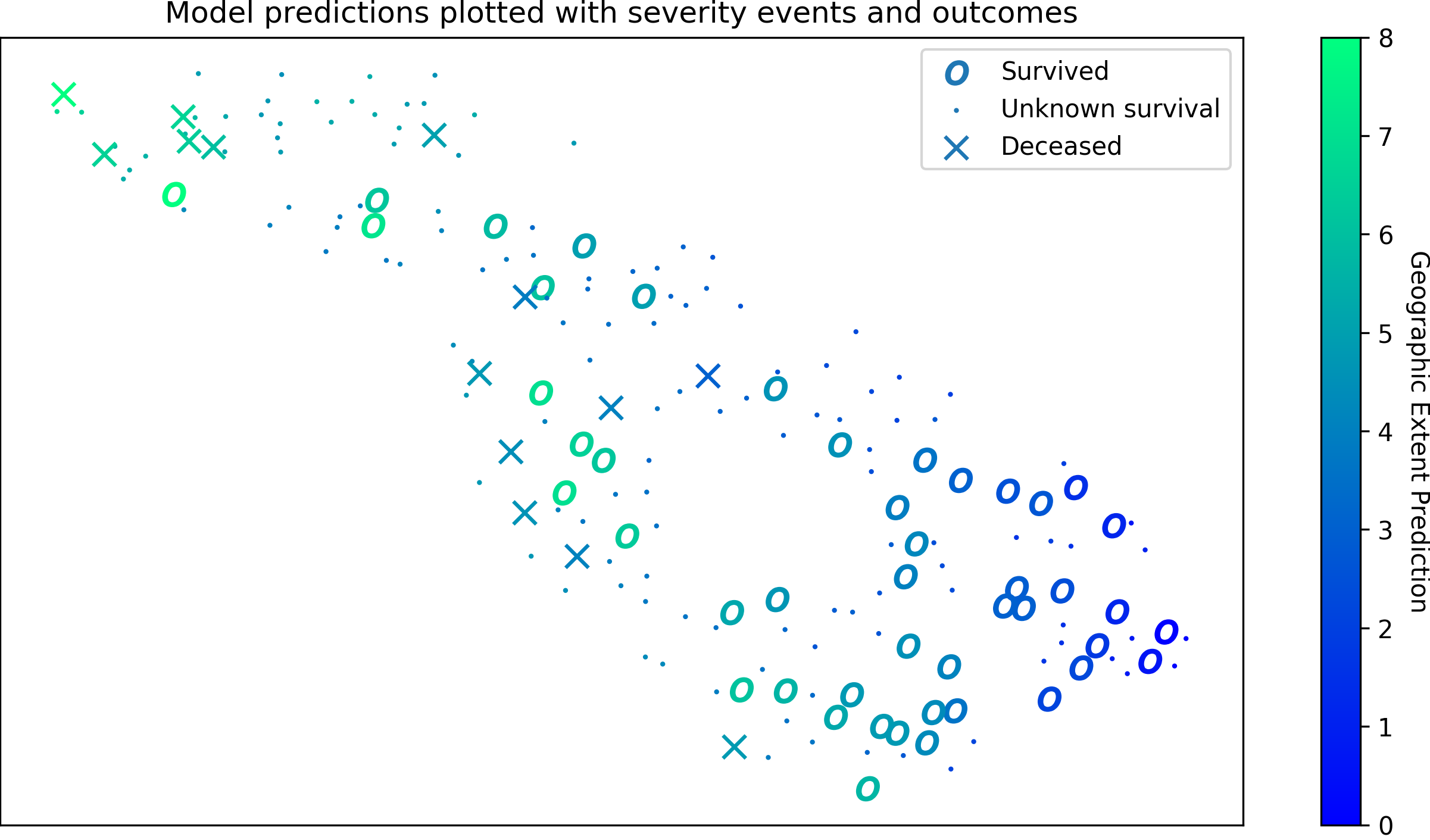}
    \vspace{-5pt}
    \caption{A spatial representation of pneumonia specific features (lung opacity, pneumonia, infiltration, and consolidation) when projected into 2 dimensions (2D) using a t-distributed stochastic neighbor embedding (t-SNE) \cite{vanDerMaaten2008_t-sne}. In this 2D space, the high dimensional (4D) distances are preserved, specifically what is nearby. CXR images which have similar outputs are close to each other. Features are extracted for all 208 images in the dataset and the geographic extent prediction is shown for each image. The survival information available in the dataset represented by the shape of the marker. }
    \label{fig:tsne-survival}
\end{figure*}
\begin{figure*}[]
\centering

    \subfigure[Geographic Extent Score: 5, Predicted: 5.3]{
    \includegraphics[width=0.24\textwidth]{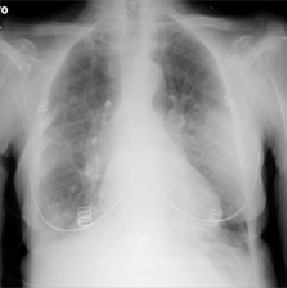}
    \includegraphics[width=0.24\textwidth]{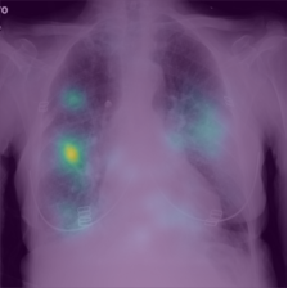}
    }%
    \subfigure[Geographic Extent Score: 0, Predicted: -0.8]{
    \includegraphics[width=0.24\textwidth]{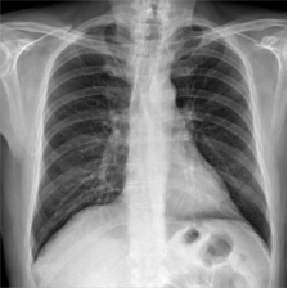}
    \includegraphics[width=0.24\textwidth]{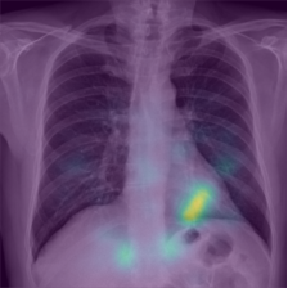}
    }

    \subfigure[Geographic Extent Score: 2, Predicted: 0.62]{
    \includegraphics[width=0.24\textwidth]{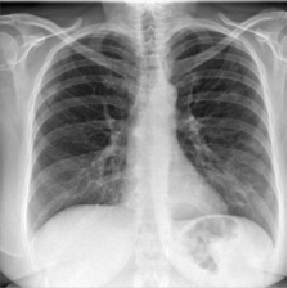}
    \includegraphics[width=0.24\textwidth]{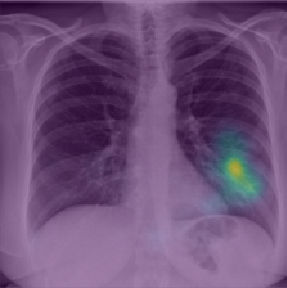}
    }%
    \subfigure[Geographic Extent Score: 0, Predicted: 1.05]{
    \includegraphics[width=0.24\textwidth]{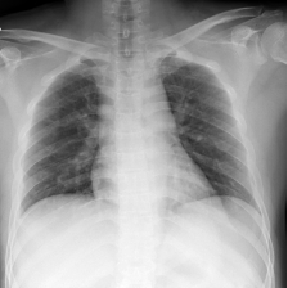}
    \includegraphics[width=0.24\textwidth]{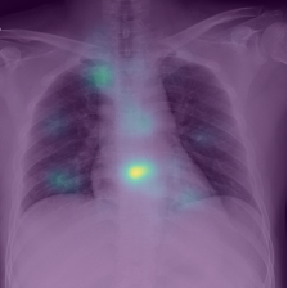}
    }
    \vspace{-5pt}
    \caption{Examples of correct (a,b) and incorrect (c,d) predictions by the model are shown with a saliency map generated by computing the gradient of the output prediction with respect to the input image and then blurred using a 5x5 Gaussian kernel. The assigned and predicted scores for Geographic Extent are shown to the right. }
    \label{fig:scatter}
\end{figure*}

\textbf{Studying learned representations} In Figure \ref{fig:tsne-survival}, we explore what the representation used by one of the best models looks at in order to identify signs of overfitting and to gain insights into the variation of the data. A t-distributed stochastic neighbor embedding (t-SNE) \cite{vanDerMaaten2008_t-sne} is computed on all data (even those not scored) in order to project the features into a two-dimensional (2D) space. Each CXR is represented by a point in a space where relationships to other points are preserved from the higher dimensional space. The cases of the survival group tend to cluster together as well as the cases of the deceased group. This clustering indicates that score predictions align with clinical outcomes.

\vspace{10pt}
\textbf{Inspecting saliency maps} In Figure \ref{fig:scatter}, images are studied which were not seen by the model during training. For most of the results, the model is correctly looking at opaque regions of the lungs. Figure \ref{fig:scatter}b shows no signs of opacity and the model is focused on the heart and diaphragm, which is likely a sign that they are used as a color reference when determining what qualifies as opaque. In Figure \ref{fig:scatter}c and \ref{fig:scatter}d, we see erroneous predictions.

\section{Discussion}

Existing work focuses on predicting severity from a variety of clinical indicators which include findings from chest imaging \cite{Jiang2020severity,Shi2020hostriskscore}. Models such as the one presented in this work can complement and improve these models and potentially help to make decisions from CXR as opposed to CT. 

Challenges in creating a predictive model involve labelling the data and achieving good inter-rater agreement as well as learning a representation which will generalize to new images when the number of labelled images is so low. In the case of building a predictive tool for COVID-19 CXR images, the lack of a public database made it difficult to conduct large-scale robust evaluations. This small number of samples prevents proper cohort selection which is a limitation of this study and exposes our evaluation to sample bias. However, we use a model which was trained on a large dataset with related tasks which provided us with a robust unbiased COVID-19 feature extractor and allows us to learn only two parameters for our best linear regression model. Restricting the complexity of the learned model in this way reduces the possibility of overfitting. 

Our evaluation could be improved if we were able to obtain new cohorts labelled with the same severity score to ascertain the generalization of our model. Also, it is unknown if these radiographic scores of disease severity reflect actual functional or clinical outcomes as the open data do not have those data. We make the images, labels, model, and code public from this work so that other groups can perform follow-up evaluations.

\section{Conclusion}

In the context of a pandemic and the urgency to contain the crisis, research has increased exponentially in order to alleviate the healthcare system’s burden. However, many prediction models for diagnosis and prognosis of COVID-19 infection are at high risk of bias and model overfitting as well as poorly reported, their alleged performance being likely optimistic \cite{Wynants2020criticalappraisal}. In order to prevent premature implementation in hospitals \cite{Ross2020}, tools must be robustly evaluated along several practical axes \cite{Wiens2019,Ghassemi2019guidance, cohen2020limits}. Indeed, while some AI-assisted tools might be powerful, they do not replace clinical judgment and their diagnostic performance cannot be assessed or claimed without a proper clinical trial \cite{Nagendran2020}.

Our model’s ability to gauge severity of COVID-19 lung infections could be used for escalation or de-escalation of care as well as monitoring treatment efficacy, especially in the intensive care unit (ICU) \cite{Toussie2020}. The use of a score combining geographical extent and degree of opacity allows clinicians to compare CXR images with each other using a quantitative and objective measure. Also, this can be done at scale for a large scale analysis.

\section*{Acknowledgements}

This research is based on work partially supported by the CIFAR AI and COVID-19 Catalyst Grants. This work utilized the supercomputing facilities managed by Compute Canada and Calcul Quebec. We thank AcademicTorrents.com for making data available for our research. 

\section*{Ethics}

This project is approved by the University of Montreal's Ethics Committee \#CERSES-20-058-D

\bibliography{refs,chester-covid-19,papers}

\begin{thebibliography}{32}
\providecommand{\natexlab}[1]{#1}
\providecommand{\url}[1]{\texttt{#1}}
\expandafter\ifx\csname urlstyle\endcsname\relax
  \providecommand{\doi}[1]{doi: #1}\else
  \providecommand{\doi}{doi: \begingroup \urlstyle{rm}\Url}\fi

\bibitem[Bustos et~al.(2019)Bustos, Pertusa, Salinas, and de~la
  Iglesia-Vay{\'{a}}]{Bustos2019PadChest}
Bustos, Aurelia, Pertusa, Antonio, Salinas, Jose-Maria, and de~la
  Iglesia-Vay{\'{a}}, Maria.
\newblock {PadChest: A large chest x-ray image dataset with multi-label
  annotated reports}.
\newblock \emph{arXiv preprint}, 1 2019.

\bibitem[Cohen et~al.(2020{\natexlab{a}})Cohen, Hashir, Brooks, and
  Bertrand]{cohen2020limits}
Cohen, Joseph~Paul, Hashir, Mohammad, Brooks, Rupert, and Bertrand, Hadrien.
\newblock {On the limits of cross-domain generalization in automated X-ray
  prediction}.
\newblock In \emph{Medical Imaging with Deep Learning}, 2020{\natexlab{a}}.

\bibitem[Cohen et~al.(2020{\natexlab{b}})Cohen, Morrison, and
  Dao]{Cohen2020coviddataset}
Cohen, Joseph~Paul, Morrison, Paul, and Dao, Lan.
\newblock {COVID-19 Image Data Collection}.
\newblock \emph{https://github.com/ieee8023/covid-chestxray-dataset},
  2020{\natexlab{b}}.

\bibitem[Cohen et~al.(2020{\natexlab{c}})Cohen, Viviano, Hashir, and
  Bertrand]{Cohen2020xrv}
Cohen, Joseph~Paul, Viviano, Joseph, Hashir, Mohammad, and Bertrand, Hadrien.
\newblock {TorchXRayVision: A library of chest X-ray datasets and models}.
\newblock \emph{https://github.com/mlmed/torchxrayvision}, 2020{\natexlab{c}}.

\bibitem[Demner-Fushman et~al.(2016)Demner-Fushman, Kohli, Rosenman, Shooshan,
  Rodriguez, Antani, Thoma, and McDonald]{Demner-Fushman2016}
Demner-Fushman, Dina, Kohli, Marc~D., Rosenman, Marc~B., Shooshan, Sonya~E.,
  Rodriguez, Laritza, Antani, Sameer, Thoma, George~R., and McDonald,
  Clement~J.
\newblock {Preparing a collection of radiology examinations for distribution
  and retrieval}.
\newblock \emph{Journal of the American Medical Informatics Association}, 3
  2016.
\newblock \doi{10.1093/jamia/ocv080}.

\bibitem[Ghassemi et~al.(2019)Ghassemi, Naumann, Schulam, Beam, Chen, and
  Ranganath]{Ghassemi2019guidance}
Ghassemi, Marzyeh, Naumann, Tristan, Schulam, P., Beam, Andrew~L., Chen,
  Irene~Y., and Ranganath, Rajesh.
\newblock {Practical guidance on artificial intelligence for health-care data},
  8 2019.

\bibitem[Huang et~al.(2017)Huang, Liu, van~der Maaten, and
  Weinberger]{Huang2017}
Huang, Gao, Liu, Zhuang, van~der Maaten, Laurens, and Weinberger, Kilian~Q.
\newblock {Densely Connected Convolutional Networks}.
\newblock In \emph{Computer Vision and Pattern Recognition}, 2017.

\bibitem[Irvin et~al.(2019)Irvin, Rajpurkar, Ko, Yu, Ciurea-Ilcus, Chute,
  Marklund, Haghgoo, Ball, Shpanskaya, Seekins, Mong, Halabi, Sandberg, Jones,
  Larson, Langlotz, Patel, Lungren, and Ng]{Irvin2019CheXpert}
Irvin, Jeremy, Rajpurkar, Pranav, Ko, Michael, Yu, Yifan, Ciurea-Ilcus,
  Silviana, Chute, Chris, Marklund, Henrik, Haghgoo, Behzad, Ball, Robyn,
  Shpanskaya, Katie, Seekins, Jayne, Mong, David~A., Halabi, Safwan~S.,
  Sandberg, Jesse~K., Jones, Ricky, Larson, David~B., Langlotz, Curtis~P.,
  Patel, Bhavik~N., Lungren, Matthew~P., and Ng, Andrew~Y.
\newblock {CheXpert: A Large Chest Radiograph Dataset with Uncertainty Labels
  and Expert Comparison}.
\newblock In \emph{AAAI Conference on Artificial Intelligence}, 1 2019.

\bibitem[Jiang et~al.(2020)Jiang, Coffee, Bari, Wang, Jiang, Huang, Shi, Dai,
  Cai, Zhang, Wu, He, and Huang]{Jiang2020severity}
Jiang, Xiangao, Coffee, Megan, Bari, Anasse, Wang, Junzhang, Jiang, Xinyue,
  Huang, Jianping, Shi, Jichan, Dai, Jianyi, Cai, Jing, Zhang, Tianxiao, Wu,
  Zhengxing, He, Guiqing, and Huang, Yitong.
\newblock {Towards an Artificial Intelligence Framework for Data-Driven
  Prediction of Coronavirus Clinical Severity}.
\newblock \emph{Computers, Materials {\&} Continua}, 2020.
\newblock \doi{10.32604/cmc.2020.010691}.

\bibitem[Jin et~al.(2020)Jin, Cai, Cheng, Cheng, Deng, Fan, Fang, Huang, Huang,
  Huang, Han, Hu, Hu, Li, Li, Liang, Lin, Luo, Ma, Ma, Peng, Pan, Pan, Ren,
  Sun, Wang, Wang, Weng, Wei, Wu, Xia, Xiong, Xu, Yao, Yuan, Ye, Zhang, Zhang,
  Zhang, Zhang, Zhao, Zhao, Zi, Zeng, Wang, and Wang]{Jin2020}
Jin, Ying-Hui, Cai, Lin, Cheng, Zhen-Shun, Cheng, Hong, Deng, Tong, Fan,
  Yi-Pin, Fang, Cheng, Huang, Di, Huang, Lu-Qi, Huang, Qiao, Han, Yong, Hu, Bo,
  Hu, Fen, Li, Bing-Hui, Li, Yi-Rong, Liang, Ke, Lin, Li-Kai, Luo, Li-Sha, Ma,
  Jing, Ma, Lin-Lu, Peng, Zhi-Yong, Pan, Yun-Bao, Pan, Zhen-Yu, Ren, Xue-Qun,
  Sun, Hui-Min, Wang, Ying, Wang, Yun-Yun, Weng, Hong, Wei, Chao-Jie, Wu,
  Dong-Fang, Xia, Jian, Xiong, Yong, Xu, Hai-Bo, Yao, Xiao-Mei, Yuan, Yu-Feng,
  Ye, Tai-Sheng, Zhang, Xiao-Chun, Zhang, Ying-Wen, Zhang, Yin-Gao, Zhang,
  Hua-Min, Zhao, Yan, Zhao, Ming-Juan, Zi, Hao, Zeng, Xian-Tao, Wang, Yong-Yan,
  and Wang, Xing-Huan.
\newblock {A rapid advice guideline for the diagnosis and treatment of 2019
  novel coronavirus (2019-nCoV) infected pneumonia (standard version).}
\newblock \emph{Military Medical Research}, 2020.
\newblock \doi{10.1186/s40779-020-0233-6}.

\bibitem[Johnson et~al.(2019)Johnson, Pollard, Berkowitz, Greenbaum, Lungren,
  Deng, Mark, and Horng]{Johnson2019mimic-cxr}
Johnson, Alistair E.~W., Pollard, Tom~J., Berkowitz, Seth~J., Greenbaum,
  Nathaniel~R., Lungren, Matthew~P., Deng, Chih-ying, Mark, Roger~G., and
  Horng, Steven.
\newblock {MIMIC-CXR: A large publicly available database of labeled chest
  radiographs}.
\newblock \emph{Nature Scientific Data}, 1 2019.
\newblock \doi{10.1038/s41597-019-0322-0}.

\bibitem[Majkowska et~al.(2019)Majkowska, Mittal, Steiner, Reicher, McKinney,
  Duggan, Eswaran, Cameron~Chen, Liu, Kalidindi, Ding, Corrado, Tse, and
  Shetty]{Majkowska2019}
Majkowska, Anna, Mittal, Sid, Steiner, David~F., Reicher, Joshua~J., McKinney,
  Scott~Mayer, Duggan, Gavin~E., Eswaran, Krish, Cameron~Chen, Po-Hsuan, Liu,
  Yun, Kalidindi, Sreenivasa~Raju, Ding, Alexander, Corrado, Greg~S., Tse,
  Daniel, and Shetty, Shravya.
\newblock {Chest Radiograph Interpretation with Deep Learning Models:
  Assessment with Radiologist-adjudicated Reference Standards and
  Population-adjusted Evaluation}.
\newblock \emph{Radiology}, 12 2019.
\newblock \doi{10.1148/radiol.2019191293}.

\bibitem[Nagendran et~al.(2020)Nagendran, Chen, Lovejoy, Gordon, Komorowski,
  Harvey, Topol, Ioannidis, Collins, and Maruthappu]{Nagendran2020}
Nagendran, Myura, Chen, Yang, Lovejoy, Christopher~A., Gordon, Anthony~C.,
  Komorowski, Matthieu, Harvey, Hugh, Topol, Eric~J., Ioannidis, John~P.A.,
  Collins, Gary~S., and Maruthappu, Mahiben.
\newblock {Artificial intelligence versus clinicians: Systematic review of
  design, reporting standards, and claims of deep learning studies in medical
  imaging}.
\newblock \emph{The BMJ}, 3 2020.
\newblock \doi{10.1136/bmj.m689}.

\bibitem[Ng et~al.(2020)Ng, Lee, Yang, Yang, Li, Wang, Lui, Lo, Leung, Khong,
  Hui, Yuen, and Kuo]{Ng2020}
Ng, Ming-Yen, Lee, Elaine Y~P, Yang, Jin, Yang, Fangfang, Li, Xia, Wang,
  Hongxia, Lui, Macy Mei-sze, Lo, Christine Shing-Yen, Leung, Barry, Khong,
  Pek-Lan, Hui, Christopher Kim-Ming, Yuen, Kwok-yung, and Kuo, Michael~David.
\newblock {Imaging Profile of the COVID-19 Infection: Radiologic Findings and
  Literature Review}.
\newblock \emph{Radiology: Cardiothoracic Imaging}, 2 2020.
\newblock \doi{10.1148/ryct.2020200034}.

\bibitem[O'Grady et~al.(2020)O'Grady, Noack, Mettler, Knowles, Armus, Wagner,
  And, and Berger]{OGrady2020}
O'Grady, Siobhán, Noack, Rick, Mettler, Katie, Knowles, Hannah, Armus, Teo,
  Wagner, John, And, Brittany~Shammas, and Berger, Miriam.
\newblock {U.S. COVID-19 death toll surpasses 2,000 in one day and 100,000
  total worldwide}, 2020.

\bibitem[Rajpurkar et~al.(2017)Rajpurkar, Irvin, Zhu, Yang, Mehta, Duan, Ding,
  Bagul, Langlotz, Shpanskaya, Lungren, and Ng]{Rajpurkar2017chexnet}
Rajpurkar, Pranav, Irvin, Jeremy, Zhu, Kaylie, Yang, Brandon, Mehta, Hershel,
  Duan, Tony, Ding, Daisy, Bagul, Aarti, Langlotz, Curtis, Shpanskaya, Katie,
  Lungren, Matthew~P., and Ng, Andrew~Y.
\newblock {CheXNet: Radiologist-Level Pneumonia Detection on Chest X-Rays with
  Deep Learning}.
\newblock \emph{arxiv}, 11 2017.

\bibitem[Reed \& Marks(1999)Reed and Marks]{Reed1999smithing}
Reed, Russell~D. and Marks, Robert~J.
\newblock \emph{{Neural smithing : supervised learning in feedforward
  artificial neural networks}}.
\newblock MIT Press, 1999.

\bibitem[Ross et~al.(2017)Ross, Hughes, and Doshi-Velez]{Ross2017rrr}
Ross, Andrew, Hughes, Michael~C, and Doshi-Velez, Finale.
\newblock {Right for the Right Reasons: Training Differentiable Models by
  Constraining their Explanations}.
\newblock In \emph{International Joint Conference on Artificial Intelligence},
  2017.

\bibitem[Ross(2020)]{Ross2020}
Ross, Casey.
\newblock {AI used to predict Covid-19 patients' decline before proven to
  work}, 2020.

\bibitem[Shi et~al.(2020)Shi, Yu, Zhao, Wang, Zhao, and
  Sheng]{Shi2020hostriskscore}
Shi, Yu, Yu, Xia, Zhao, Hong, Wang, Hao, Zhao, Ruihong, and Sheng, Jifang.
\newblock {Host susceptibility to severe COVID-19 and establishment of a host
  risk score: Findings of 487 cases outside Wuhan}.
\newblock \emph{Critical Care}, 12 2020.
\newblock \doi{10.1186/s13054-020-2833-7}.

\bibitem[Shih et~al.(2019)Shih, Wu, Halabi, Kohli, Prevedello, Cook, Sharma,
  Amorosa, Arteaga, Galperin-Aizenberg, Gill, Godoy, Hobbs, Jeudy, Laroia,
  Shah, Vummidi, Yaddanapudi, and Stein]{Shih2019RSNAKaggle}
Shih, George, Wu, Carol~C., Halabi, Safwan~S., Kohli, Marc~D., Prevedello,
  Luciano~M., Cook, Tessa~S., Sharma, Arjun, Amorosa, Judith~K., Arteaga,
  Veronica, Galperin-Aizenberg, Maya, Gill, Ritu~R., Godoy, Myrna~C.B., Hobbs,
  Stephen, Jeudy, Jean, Laroia, Archana, Shah, Palmi~N., Vummidi, Dharshan,
  Yaddanapudi, Kavitha, and Stein, Anouk.
\newblock {Augmenting the National Institutes of Health Chest Radiograph
  Dataset with Expert Annotations of Possible Pneumonia}.
\newblock \emph{Radiology: Artificial Intelligence}, 1 2019.
\newblock \doi{10.1148/ryai.2019180041}.

\bibitem[Strickland(2020)]{Strickland2020}
Strickland, Eliza.
\newblock {AI Can Help Hospitals Triage COVID-19 Patients}, 2020.

\bibitem[Toussie et~al.(2020)Toussie, Voutsinas, Finkelstein, Cedillo, Manna,
  Maron, Jacobi, Chung, Bernheim, Eber, Concepcion, Fayad, and
  Gupta]{Toussie2020}
Toussie, Danielle, Voutsinas, Nicholas, Finkelstein, Mark, Cedillo, Mario~A,
  Manna, Sayan, Maron, Samuel~Z, Jacobi, Adam, Chung, Michael, Bernheim, Adam,
  Eber, Corey, Concepcion, Jose, Fayad, Zahi, and Gupta, Yogesh~Sean.
\newblock {Clinical and Chest Radiography Features Determine Patient Outcomes
  In Young and Middle Age Adults with COVID-19}.
\newblock \emph{Radiology}, 5 2020.
\newblock \doi{10.1148/radiol.2020201754}.

\bibitem[van~der Maaten \& Hinton(2008)van~der Maaten and
  Hinton]{vanDerMaaten2008_t-sne}
van~der Maaten, Laurens and Hinton, Geoffrey.
\newblock {Visualizing Data using t-SNE}.
\newblock \emph{Journal of Machine Learning Research}, 2008.

\bibitem[Viviano et~al.(2019)Viviano, Simpson, Dutil, Bengio, and
  Cohen]{Viviano2019}
Viviano, Joseph~D., Simpson, Becks, Dutil, Francis, Bengio, Yoshua, and Cohen,
  Joseph~Paul.
\newblock {Underwhelming Generalization Improvements From Controlling Feature
  Attribution}.
\newblock \emph{arxiv:1910.00199}, 10 2019.

\bibitem[Wang et~al.(2017)Wang, Peng, Lu, Lu, Bagheri, and
  Summers]{WangNIH2017}
Wang, Xiaosong, Peng, Yifan, Lu, Le, Lu, Zhiyong, Bagheri, Mohammadhadi, and
  Summers, Ronald~M.
\newblock {ChestX-ray8: Hospital-scale Chest X-ray Database and Benchmarks on
  Weakly-Supervised Classification and Localization of Common Thorax Diseases}.
\newblock In \emph{Computer Vision and Pattern Recognition}, 2017.
\newblock \doi{10.1109/CVPR.2017.369}.

\bibitem[Wiens et~al.(2019)Wiens, Saria, Sendak, Ghassemi, Liu, Doshi-Velez,
  Jung, Heller, Kale, Saeed, Ossorio, Thadaney-Israni, and
  Goldenberg]{Wiens2019}
Wiens, Jenna, Saria, Suchi, Sendak, Mark, Ghassemi, Marzyeh, Liu, Vincent~X.,
  Doshi-Velez, Finale, Jung, Kenneth, Heller, Katherine, Kale, David, Saeed,
  Mohammed, Ossorio, Pilar~N., Thadaney-Israni, Sonoo, and Goldenberg, Anna.
\newblock {Do no harm: a roadmap for responsible machine learning for health
  care}.
\newblock \emph{Nature Medicine}, 8 2019.
\newblock \doi{10.1038/s41591-019-0548-6}.

\bibitem[Wilson \& Moulson(2020)Wilson and Moulson]{Wilson2020}
Wilson, Joseph and Moulson, Geir.
\newblock {Children in Spain allowed to play outdoors as country eases COVID-19
  lockdown}, 2020.

\bibitem[Wong et~al.(2019)Wong, Lam, Fong, Leung, Chin, Lo, Lui, Lee, Chiu,
  Chung, Lee, Wan, Hung, Lam, Kuo, and Ng]{10.1148/radiol.2020201160}
Wong, Ho Yuen~Frank, Lam, Hiu Yin~Sonia, Fong, Ambrose Ho~Tung, Leung,
  Siu~Ting, Chin, Thomas Wing~Yan, Lo, Christine Shing~Yen, Lui, Macy Mei~Sze,
  Lee, Jonan Chun~Yin, Chiu, Keith Wan~Hang, Chung, Tom, Lee, Elaine Yuen~Phin,
  Wan, Eric Yuk~Fai, Hung, Fan Ngai~Ivan, Lam, Tina Poy~Wing, Kuo, Michael, and
  Ng, Ming~Yen.
\newblock {Frequency and Distribution of Chest Radiographic Findings in
  COVID-19 Positive Patients}.
\newblock \emph{Radiology}, 3 2019.
\newblock \doi{10.1148/radiol.2020201160}.

\bibitem[Wynants et~al.(2020)Wynants, Van~Calster, Bonten, Collins, Debray,
  De~Vos, Haller, Heinze, Moons, Riley, Schuit, Smits, Snell, Steyerberg,
  Wallisch, and Van~Smeden]{Wynants2020criticalappraisal}
Wynants, Laure, Van~Calster, Ben, Bonten, Marc~M.J., Collins, Gary~S., Debray,
  Thomas~P.A., De~Vos, Maarten, Haller, Maria~C., Heinze, Georg, Moons,
  Karel~G.M., Riley, Richard~D., Schuit, Ewoud, Smits, Luc~J.M., Snell,
  Kym~I.E., Steyerberg, Ewout~W., Wallisch, Christine, and Van~Smeden, Maarten.
\newblock {Prediction models for diagnosis and prognosis of COVID-19 infection:
  Systematic review and critical appraisal}.
\newblock \emph{The BMJ}, 4 2020.
\newblock \doi{10.1136/bmj.m1328}.

\bibitem[Yoon et~al.(2020)Yoon, Lee, Kim, Lee, Ko, Kim, Park, and
  Kim]{Yoon2020}
Yoon, Soon~Ho, Lee, Kyung~Hee, Kim, Jin~Yong, Lee, Young~Kyung, Ko, Hongseok,
  Kim, Ki~Hwan, Park, Chang~Min, and Kim, Yun-Hyeon.
\newblock {Chest Radiographic and CT Findings of the 2019 Novel Coronavirus
  Disease (COVID-19): Analysis of Nine Patients Treated in Korea}.
\newblock \emph{Korean Journal of Radiology}, 2020.
\newblock \doi{10.3348/kjr.2020.0132}.

\bibitem[Zech et~al.(2018)Zech, Badgeley, Liu, Costa, Titano, and
  Oermann]{Zech2018}
Zech, John~R., Badgeley, Marcus~A., Liu, Manway, Costa, Anthony~B., Titano,
  Joseph~J., and Oermann, Eric~Karl.
\newblock {Variable generalization performance of a deep learning model to
  detect pneumonia in chest radiographs: A cross-sectional study}.
\newblock \emph{PLoS Medicine}, 7 2018.
\newblock \doi{10.1371/journal.pmed.1002683}.

\end{thebibliography}
\bibliographystyle{icml-nopagenum}

\end{document}